\documentclass[fleqn,twoside]{article}
\usepackage[headings]{espcrc2}
\usepackage{graphicx}
\usepackage{booktabs}
\usepackage{multirow}
\usepackage{subfig, endnotes, }
\usepackage{algorithm}
\usepackage{algorithmic}
\usepackage{amssymb}
\usepackage{array,colortbl,xcolor}
\usepackage{tikz}
\usetikzlibrary{backgrounds}
\usepackage{url}
\readRCS
$Id: espcrc2.tex,v 1.2 2004/02/24 11:22:11 spepping Exp $
\usepackage{graphicx, balance}
\usepackage[figuresright]{rotating}

\newcommand{\AmS}{{\protect\the\textfont2
  A\kern-.1667em\lower.5ex\hbox{M}\kern-.125emS}}

\title{\textbf{Use Case Point Approach Based Software Effort Estimation using Various Support Vector Regression Kernel Methods}}

\author{Shashank Mouli Satapathy\address[DCSE]{Department of Computer Science and Engineering, National Insitutte of Technology Rourkela, Rourkela 769008 India, Contact: shashankamouli@gmail.com \\},
Santanu Kumar Rath\address{Department of Computer Science and Engineering, National Insitutte of Technology Rourkela, Rourkela 769008 India, Contact: skrath@nitrkl.ac.in}}

\runtitle{UCP Based Software Effort Estimation using Various SVR Kernel Methods}
\runauthor{Satapathy S M, et al.,}

\begin{document}
\begin{abstract}

The job of software effort estimation is a critical one in the early stages of the software development life cycle when the details of requirements are usually not clearly identified. Various optimization techniques help in improving the accuracy of  effort estimation. The Support Vector Regression (SVR) is one of several different soft-computing techniques that help in getting optimal estimated values. The idea of SVR is based upon the computation of a linear regression function in a high dimensional feature space where the input data are mapped via a nonlinear function.  Further, the SVR kernel methods can be applied in transforming the input data and then based on these transformations, an optimal boundary between the possible outputs can be obtained. The main objective of the research work carried out in this paper is to estimate the software effort using use case point approach. The use case point approach relies on the use case diagram to estimate the size and effort of software projects. Then, an attempt has been made to optimize the results obtained from use case point analysis using various SVR kernel methods to achieve better prediction accuracy. 
\end{abstract}

\maketitle

\section{INTRODUCTION}
Proper software effort estimation is the foremost activity adopted in every software development life cycle. Several features offered by OO programming concept such as Encapsulation, Inheritance, Polymorphism, Abstraction, Cohesion and Coupling play an important role to manage the development process \cite{Carbone,Gill}. Currently used software development effort estimation models such as, COCOMO and Function Point Analysis (FPA), do not consistently provide accurate project cost and effort estimates \cite{Xu}. These techniques have been proven unsatisfactory for estimating cost and effort because the lines of code (LOC) and function point (FP) are both used for procedural oriented paradigm \cite{Matson}. Both of them have certain limitations. The LOC is dependent on the programming language and the FPA is based on human decisions. Hence effort estimation during early stage of software development life cycle plays a vital role for determining whether a project is feasible in terms of a cost-benefit analysis \cite{MacDonell,Strike}.

The Use Case Point (UCP) model relies on the use case diagram to estimate the effort of a given software product. UCP helps in providing more accurate effort estimation from design phase of software development life cycle. UCP is measured by counting the number of use cases and the number of actors, each multiplied by its complexity factors. Use cases and actors are classified into three categories. These include simple, average and complex. The determination of the complexity value (simple, average or complex) of use cases is determined by the number of transactions per use case. The UCP model has widely been used in the last decade \cite{Damodaran}, yet it has several limitations. One of the limitations is that the software effort equation is not well accepted by software estimators because it assumes that the relationship between software effort and size is linear. In this paper, various kernel methods based support vector regression is introduced to tackle the limitations of the UCP model and to enhance prediction accuracy of the software effort estimation. The result obtained from these support vector regression techniques based effort estimation model is then compared with other models in order to assess performance of those models.

\section{RELATED WORK}
This section presents related work regarding the use case point based effort estimation approach and support vector regression model.

A. Issha et al., \cite{Issa} reports on the development of three novel use case model based software cost estimation methods such as use case rough estimation method, use case patterns estimation method, and object points extraction estimation method. The accuracy of the proposed methods has been investigated using a wide spectrum of software projects. Ali B. Nassif et al., \cite{Nassif2011Regression} presents a novel regression model to estimate the software effort based on the use case point size metric. They proposed an effort equation that takes into consideration the non-linear relationship between software size and software effort, as well as the influences of project complexity and productivity. Results show that the software effort estimation accuracy can be improved by 16.5\%. Ali B. Nassif et al., \cite{Nassif2011rRegressionMFIS} extended this process by applying mamdani fuzzy inference system with regression model to enhance the estimation accuracy and found 10\% improvement in the result. Ali B. Nassif et al., \cite{Nassif2011RegressionSFIS} also applied  sugeno fuzzy inference system with regression model to enhance the estimation accuracy and found 11\% improvement in the MMRE result. Ali B. Nassif et al., \cite{Nassif2012ANN} propose a novel Artificial Neural Network (ANN) to predict software effort from use case diagrams based on the Use Case Point (UCP) model with the help of 240 data points and found a competitive result with respect to other regression model. Ali B. Nassif et al., \cite{Nassif2010ANNFuzzy} also present some techniques using fuzzy logic and neural networks to improve the accuracy of the use case points method and obtained an improvement up to 22\%. Ali B. Nassif et al., \cite{Nassif2012Treeboost} uses a Treeboost (Stochastic Gradient Boosting) model to predict software effort based on the Use Case Point method using 84 data points and obtain promising results. 

Adriano L.I. Oliveira \cite{Oliveira} provides a comparative study on support vector regression (SVR), radial basis function neural networks (RBFNs) and linear regression for estimation of software project effort. The experiment is carried out using NASA project datasets and the result shows that SVR performs better than RBFN and linear regression. Petr{\^o}nio L. Braga et al., \cite{Braga} have proposed and investigated the use of a genetic algorithm approach for selecting an optimal feature subset and optimizing SVR parameters simultaneously aiming to improve the precision of the software effort estimates. E. Kocaguneli et al., \cite{Kocaguneli} have investigated non-uniform weighting through kernel density estimation and found that nonuniform weighting through kernel methods cannot outperform uniform weighting Analogy Based Estimation (ABE). Bilge Başkeleş et al., \cite{Baskeles2007} propose a model that uses machine learning methods and evaluate the model on public data sets and data gathered from software organization. From analysis, it is found out that the usage of any one model cannot produce the best results for software effort estimation.

\section{METHODOLOGY USED}
The following methodologies are used in this paper to calculate the effort of a software product.

\subsection{Use Case Point Approach}

\begin{figure}[ht!]
\centering
\begin{tikzpicture}[scale=1]

\node[text width=1.7cm, text badly centered,draw,fill=none, very thick, font=\scriptsize] at (1,9.5) (a) {\textbf{Use Case Diagram}};

\node[text width=1.7cm, text badly centered,draw,fill=none, very thick, font=\scriptsize] at (1,7.5) (b) {\textbf{Classification of Actors and Use Cases}};  
                                                      
\node[text width=1.7cm, text badly centered,draw,fill=none, very thick, font=\scriptsize] at (3.8,7.5) (c) {\textbf{Calculation of Weights and Points}};  
                                              
\node[text width=1.7cm, text badly centered,draw,fill=none, very thick, font=\scriptsize] at (6.5,7.5) (d) {\textbf{Calculation of  TCF and EF}};  

\node[text width=1.7cm, text badly centered,draw,fill=none, very thick, font=\scriptsize] at (6.5,5.5) (e) {\textbf{Final Use Case Point Evaluation}}; 
\draw[->, very thick] (a) to (b);
\draw[->, very thick] (b) to (c);
\draw[->, very thick] (c) to (d);
\draw[->, very thick] (d) to (e);
\end{tikzpicture}
\caption{Steps to Calculate Use Case Point}
 \label{figStepUCP}
\end{figure}
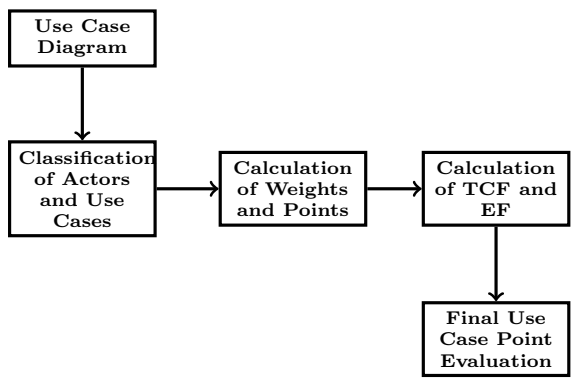

The Use Case Point (UCP) model was proposed by Gustav Karner in 1993 \cite{Karner}. This method is an extension of Function Point Analysis and Mk II Function Point Analysis (an adaptation of FPA mainly used in the UK), and is based on the same philosophy as these methods. An early estimate of effort based on use cases can be made when there is some understanding of the problem domain, system size and architecture at the stage at which the estimate is made. The block diagram shown in Figure~\ref{figStepUCP} explains the steps to calculate the class point.

The use case point approach can be implemented using the following steps:
\subsubsection{Classification of Actors and Use Cases}
The  first step is to classify the actors as simple, average or complex. A simple actor represents another system with a defined Application Programming Interface, API, an average actor is another system interacting through a protocol such as TCP/IP, and a complex actor may be a person interacting through a  GUI or a Web page. A weighting factor is assigned to each actor type in the following manner:

\begin{table}[htb!]
  \centering
  \caption{Actor Weighting Factors}
  \renewcommand{\arraystretch}{1.25}
  \resizebox{7cm}{!}{
    \begin{tabular}{|c|c|}
    \hline
    \multicolumn{1}{|c|}{\textbf{Actor Type}} & \multicolumn{1}{c|}{\textbf{Weighting Factor}} \\
    \hline
    Simple & 1 \\ \hline
    Average & 2 \\ \hline
    Complex & 3 \\ \hline
    \end{tabular}%
    }
  \label{weightActorType}%
\end{table}%

Similarly each use case is defined as simple, average or complex, depending on number of transactions in the use case description, including secondary scenarios. A transaction is a set of activities, which is either performed entirely, or not at all. Counting number of transactions can be done by counting the use case steps. The use case is considered as \textit{Simple}, when it uses a simple user interface and touches only a single database entity. The use case is considered as \textit{Average}, when it involves more interface design and touches 2 or more database entities. Similarly the use case is \textit{Complex}, when it involves a complex  user interface or processing and  touches 3 or more database entities. Use case complexity is then defined and weighted in the  following manner: 

\begin{table}[htb!]
  \centering
  \caption{Use Case Weighting Factors}
  \renewcommand{\arraystretch}{2}
  \resizebox{7cm}{!}{
    \begin{tabular}{|c|c|c|}
    \hline
    \multicolumn{1}{|c|}{\textbf{Use Case Type}} & \multicolumn{1}{|c|}{\textbf{No. of Transactions Type}} & \multicolumn{1}{c|}{\textbf{Weighting Factor}} \\
    \hline
    Simple & $<$= 3 & 5 \\ \hline
    Average & 4 to 7 & 10 \\ \hline
    Complex & $>$= 7 & 15 \\ \hline
    \end{tabular}%
    }
  \label{weightUsecaseType}%
\end{table}%

\subsubsection{Calculation of Weights and Points}
The total Unadjusted Actor Weights (UAW) is calculated by counting how many actors there are of each kind (by degree of complexity), multiplying each total by its weighting factor, and adding up the products. Similarly each type of use case is then multiplied by the weighting factor, and the products are added up to get the Unadjusted Use Case Weights (UUCW). Then the UAW is added to the UUCW to get the Unadjusted Use Case Points (UUCP).
\begin{equation} \label{UUCPCalc}
UUCP = UAW + UUCW
\end{equation}

\subsubsection{Calculation of TCF and EF}
The UUCP are adjusted based on the values`assigned to a number of technical and environmental factors shown in Tables \ref{TechFacs} and \ref{EnvFacs}. Each factor is assigned a value between 0 and 5 depending on its assumed influence on the project. A rating of 0 means the factor is irrelevant for this project and 5 means it is essential.

\begin{table}[htb!]
  \centering
  \caption{Technical Factors}
  \renewcommand{\arraystretch}{1.4}
  \resizebox{7cm}{!}{
    \begin{tabular}{|c|c|c|}
    \hline
    \multicolumn{1}{|c|}{\textbf{Factor}} & \multicolumn{1}{|c|}{\textbf{Description}} & \multicolumn{1}{c|}{\textbf{Weight}} \\
    \hline
    T1 & Distributed System & 2 \\ \hline
    T2 & Response Adjectives & 2 \\ \hline
    T3 & End-user Efficiency & 1 \\ \hline
    T4 & Complex Processing & 1 \\ \hline
    T5 & Reusable Code & 1 \\ \hline
    T6 & Easy to Install & 0.5 \\ \hline
    T7 & Easy to Use & 0.5 \\ \hline
    T8 & Portable & 2 \\ \hline
    T9 & Easy to Change & 1 \\ \hline
    T10 & Concurrent & 1 \\ \hline
    T11 & Security Features & 1 \\ \hline
    T12 & Access for Third Parties & 1 \\ \hline
    T13 & Special Training Required & 1 \\ \hline    
    \end{tabular}%
    }
  \label{TechFacs}%
\end{table}%

\begin{table}[htb!]
  \centering
  \caption{Environment Factors}
  \renewcommand{\arraystretch}{1.4}
  \resizebox{7cm}{!}{
    \begin{tabular}{|c|c|c|}
    \hline
    \multicolumn{1}{|c|}{\textbf{Factor}} & \multicolumn{1}{|c|}{\textbf{Description}} & \multicolumn{1}{c|}{\textbf{Weight}} \\
    \hline
    F1 & Familiar with RUP & 1.5 \\ \hline
    F2 & Application Experience & 0.5 \\ \hline
    F3 & Object-oriented Experience & 1 \\ \hline
    F4 & Lead Analyst Capability & 0.5 \\ \hline
    F5 & Motivation & 1 \\ \hline
    F6 & Stable Requirements & 2 \\ \hline
    F7 & Part-time Workers & -1 \\ \hline
    F8 & Difficult Programming Language & 2 \\ \hline
    \end{tabular}%
    }
  \label{EnvFacs}%
\end{table}%

The adjustment factors are multiplied by the unadjusted use case points to produce the adjusted use case points, yielding an estimate of the size of the software. The Technical Complexity Factor (TCF) is calculated by multiplying the value of each factor (T1- T13) by its weight and then adding all these numbers to get the sum called the \textit{TFactor}. The following formula is applied to find TCF: 
\begin{equation} \label{TCFcalc}
TCF = 0.6+(0.01*TFactor)
\end{equation}

The Environmental Factor (EF) is calculated by multiplying the value of each factor (F1-F8) by its weight and adding the products to get the sum called the \textit{EFactor}. The following equation gives EF value: 
\begin{equation} \label{EFcalc}
EF = 1.4+(-0.03*EFactor)
\end{equation}

\subsubsection{Final Use Case Point Calculation}
The final adjusted use case points (UCP) are calculated as follows:
\begin{equation} \label{UCPcalc}
UCP = UUCP * TCF * EF 
\end{equation}

The final use case point value is then taken as input argument to support vector regression model to calculate estimated normalized effort. 

\subsection{Support Vector Regression Technique}
Support Vector Machines (SVM) are learning machines implementing the structural risk minimization inductive principle to obtain a good generalization on a limited number of learning patterns. A version of SVM for regression was proposed by Vladimir N. Vapnik et al. \cite{Drucker} in 1996. This method is called as \textit{support vector regression (SVR)}. Generally any neural networks suffers from two major drawbacks. First of all neural networks often converge on local minima rather than global minima. Secondly neural networks often over-fit which means, if training on a pattern goes on too long, then it may consider noise as part of pattern. SVR doesn’t suffer from either of these two drawbacks and have the advantages due to which it can be successfully used for regression task.

Suppose, for a given training data {($x_{1}$, $y_{1}$), . . . , $(x_{l}$, $y_{l}$)}, where $x \in \mathbb{R}^{n}$ denotes the space of the input patterns and $y \in \mathbb{R}$ denotes its corresponding target value. The goal of regression is to find the function $f(x)$ that best models the training data. For the case of nonlinear regression, $f(x) = \langle w, \phi (x)\rangle + b$, where $\phi$ is a nonlinear function which maps the input space to a higher (maybe infinite) dimensional feature space and $\langle.,.\rangle$ denotes the dot product in $\mathbb{R}^{n}$. In SVR, the weight vector `$w$' and the threshold `$b$' are chosen to optimize the following problem \cite{Smola}.

\begin{equation}\label{eq:svrsoln}
  \begin{array}{>{$}p{.41\textwidth}<{$}}
   \min_{w,b,\xi, \xi^{*}}  = (1/2w^{T}w + C\sum_{i=1}^{l}(\xi_{i} + \xi^{*}_{i}))  \\ \\
   subject~to~~\\
   \hspace*{1cm} (\langle w, \phi (x_{i}) \rangle + b) - y_{i} \leq \epsilon + \xi_{i},\\
   \hspace*{1cm}  y_{i} - (\langle w, \phi (x_{i}) \rangle + b) +
  \leq \epsilon + \xi_{i}^{*},\\
   \hspace*{1cm}  \xi_{i}, \xi_{i}^{*} \geq 0.
  \end{array}
\end{equation}
$C > 0$ is the \textit{penalty parameter} of the error term. $\xi$ and $\xi^{*}$ are called \textit{slack variables} and measure the cost of the errors on the training points. $\xi$ measures deviations exceeding the target value by more than $\epsilon$ and $\xi^{*}$ measures deviations which are more than $\epsilon$, but below the target value \cite{Oliveira}.

$K(x_{i}, x_{j}) = \phi(x_{i}^{T}\phi(x_{j}))$ is called the \textit{kernel function}. There are basically four kernels. These are:
\begin{itemize}
\itemsep0em
\item \textbf{linear:}\\
\hspace*{0.25cm}$K(x_{i}, x_{j}) = x_{i}^{T}x_{j}$.
\item \textbf{polynomial:}\\
\hspace*{0.25cm}$K(x_{i}, x_{j}) = (\gamma x_{i}^{T}x_{j} + r)^{d}, \gamma > 0$.
\item \textbf{Radial Basis Function (RBF):}\\
\hspace*{0.25cm}$K(x_{i}, x_{j}) = \exp(-\gamma \left \| x_{i} - x_{j} \right \|^{2}), \gamma > 0$.
\item \textbf{sigmoid:}\\
\hspace*{0.25cm}$K(x_{i}, x_{j}) = tanh(\gamma x_{i}^{T}x_{j} + r)$.
\end{itemize}
Here $\gamma$, $r$, and $d$ are \textit{kernel parameters}. In epsilon-SV regression \cite{Smola}, the goal is to find a function $f(x)$ that has at most $\epsilon$ deviation from the actually obtained targets $y_{i}$ for all the training data, and at the same time is as flat as possible. In other words, errors less than $\epsilon$ are ignored and considered as zero. But errors larger than $\epsilon$ are measured by variable $\xi$ and $\xi^{*}$. The following tunable parameters \cite{Chang} have been used while implementing support vector regression.
\begin{itemize}
\itemsep0em
\item \textbf{param}: This is a string which specifies the model parameters. For regression model, a typical parameter string may look like, \textit{`-s 3 -t 2 -c 20 -g 64 -p 1'} where
\begin{itemize}
\itemsep0em
\item \textbf{-s}: svm type, 
\item \textbf{-t}: kernel type
\item \textbf{-c}: penalty parameter C of epsilon-SV regression.
\item \textbf{-g}: width parameter $\gamma$ 
\item \textbf{-p}: $\epsilon$ for epsilon-SV regression.
\end{itemize}
\end{itemize}
The value of parameter `\textit{s}' ranges from 0 to 5 and the default value is 0. For epsilon-SV regression, the parameter `\textit{s}' is assigned with value 3. The `\textit{t}' value ranges from 0 to 3 for different types of kernel. In this case, the value can be 0, 1, 2 or 3 for linear, polynomial, RBF and sigmoid kernel respectively. The default value for `\textit{t}' is 2. Similarly, the value of parameter `\textit{c}' will be calculated as the difference between maximum and minimum value of actual effort used to train the model. The default value is 1. The parameter `\textit{g}' value shows width parameter i.e., it set $\gamma$ in various kernel function. The default value is 1. Lastly, the value of parameter `\textit{p}' set the $\epsilon$ in loss function of epsilon-SVR. The default value for parameter `\textit{p}' is 0.1. The use of these tunable parameters helps in accurately estimating the predicted effort using various SVR kernel methods.

\section{PROPOSED APPROACH}
\begin{figure}[ht!]
\centering
\begin{tikzpicture}[scale=0.97, show background rectangle]

 
\node[text width=3.7cm, text badly centered,draw,fill=none, very thick, font=\scriptsize] at (1,9.5) (a) {\textbf{Calculation of Use Case Point}};

\node[text width=3.7cm, text badly centered,draw,fill=none, very thick, font=\scriptsize] at (1.5,8.2) (b) {\textbf{Scaling of Data Set}}; 
                                                       
\node[text width=3.7cm, text badly centered,draw,fill=none, very thick, font=\scriptsize] at (2,7) (c) {\textbf{Division of Data Set}};                                          

\node[text width=3.7cm, text badly centered,draw,fill=none, very thick, font=\scriptsize] at (2.5,5.8) (d) {\textbf{Performing Parameters \& Model Selection}};  

\node[text width=3.7cm, text badly centered,draw,fill=none, very thick, font=\scriptsize] at (3,4.4) (e) {\textbf{Selection of Parameters}}; 

\node[text width=3.7cm, text badly centered,draw,fill=none, very thick, font=\scriptsize] at (3.5,3.2) (f) {\textbf{Selection of Best Model}}; 

\node[text width=3.7cm, text badly centered,draw,fill=none, very thick, font=\scriptsize] at (4,1.8) (g) {\textbf{Training of Selected Model}};

\node[text width=3.7cm, text badly centered,draw,fill=none, very thick, font=\scriptsize] at (4.5,0.4) (h) {\textbf{Calculation of Error and Prediction Accuracy}}; 

\draw[->, very thick] (a) to (b);
\draw[->, very thick] (b) to (c);
\draw[->, very thick] (c) to (d);
\draw[->, very thick] (d) to (e);
\draw[->, very thick] (e) to (f);
\draw[->, very thick] (f) to (g);
\draw[->, very thick] (g) to (h);
\end{tikzpicture}
\caption{Proposed Steps Used for Effort Estimation using Various SVR Kernel Methods }
 \label{propstep}
\end{figure}
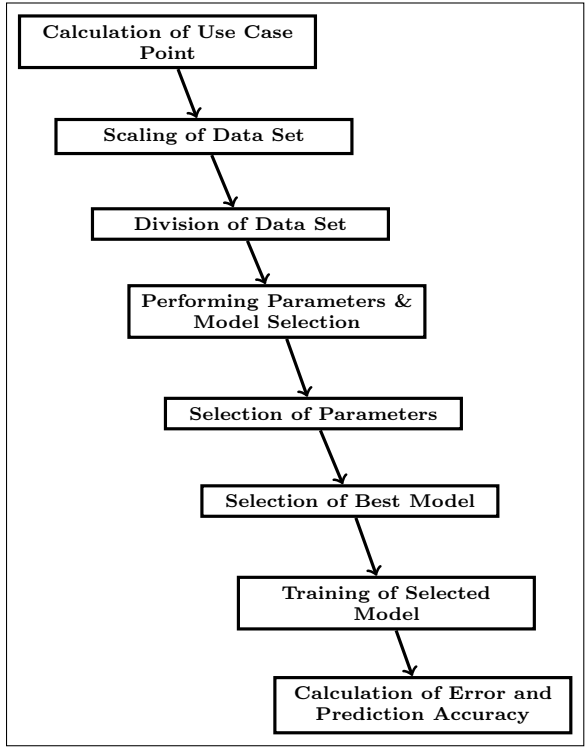

The proposed approach is based on eight-four data set used in the article \cite{Nassif2012Treeboost}. The use of this data set intends to evaluate software development effort and validate the practicability of improvement. The use of such data in the validation process has provided initial experimental evidence of the effectiveness of the UCP. These data are used to develop the support vector regression based software effort estimation model. The block diagram shown in Figure~\ref{propstep} displays the proposed steps used to determine the predicted effort using various kernel based support vector regression technique.

To calculate the effort of a given software project, basically the following steps have been used.\\[0.1cm]
\textbf{\underline{Steps in Effort Estimation}}
\begin{enumerate}
\itemsep0em
\item \textbf{Calculation of Use Case Point}: After collecting the data from previously developed projects, the use case point (UCP) has been calculated from the use case diagram using the steps proposed by Karner \cite{Karner}. 
\item \textbf{Scaling of Data Set}: The generated UCP value in \textit{Step-1} has been used as input arguments and has been scaled within the range [0,1]. Let X be the dataset and x is an element of the dataset, then the scaled value of x can be calculated as :
\begin{equation} \label{eq:normalization}
x' = \frac{x - \min(X)}{\max(X) - \min(X)}
\end{equation}

where\\
$x'$ = Scaled value of x within the range [0,1].\\
$\min(X)$ = the minimum value for the dataset X.\\
$\max(X)$ = the maximum value for the dataset X.\\
If $\max(X)$ is equal to $\min(X)$, then Normalized(x) is set to 0.5. 
\item \textbf{Division of Data Set}: The data set is divided into two parts i.e., training set \& test set. The training set is used for model estimation, whereas the test set is used only for estimating the predicted effort of the final model. While selecting a model of optimal complexity, divide the training set into a two parts i.e., learning set \& validation set. The learning set is used to estimate model parameters, whereas the validation set is used for selecting an optimal model complexity. This step is implemented using 5-fold cross validation.
\item \textbf{Performing Parameters \& Model Selection}: The model which provides the least value than the other generated models based on the minimum validation error criteria has been selected to perform other operations.
\item \textbf{Selection of Parameters}: The tunable parameters are selected to find the best value of C and $\gamma$ using a five fold cross-validation procedure .
\item \textbf{Selection of Best Model}: Based on the minimum validation error, the best model has been selected and the corresponding value of $\gamma$ and $\epsilon$ value is found out.
\item \textbf{Training of Selected Model}: The final model selected based on best parameter of C, $\epsilon$ and $\gamma$ are trained using all training samples. The output of this step is the trained SVM model providing predicted response values for test inputs.
\item \textbf{Calculation of Error and Prediction Accuracy}: The performance of the model is checked by calculating Root Mean Square Error (RMSE), Mean Magnitude of Relative Error (MMRE) and Prediction Accuracy (PRED) for test sample and training sample. The graphs generated using obtained results indicate visual comparison of the parameters.
\end{enumerate}

The above steps are followed to implement the SVR based effort estimation model using various kernel methods. Finally, a comparison of results obtained using various kernels based SVR effort estimation model has been presented to assess their performances.

\section{EXPERIMENTAL DETAILS}

In this paper to implement the proposed approaches, eighty four dataset is being used which is also used by Ali B. Nasif et al. \cite{Nassif2012Treeboost}. The detail description about the data set has already been provided in proposed approach section. The data set is divided into different subsets for the training, testing and validation purpose. First of all, every fifth data out of those two data sets, is extracted for testing  purpose and rest data will be used for training purpose. Then the training data has been partitioned into the learning / validation sets. The 5-fold data partitioning is done on the training data by the following strategy :\\[0.2cm]
\textbf{For partition 1:} Samples 1, 6, 11, ... have been used as validation samples and the remaining as learning samples\\[0.1cm]
\textbf{For partition 2:} Samples 2, 7, 12, ... have been used as validation samples and the remaining as learning samples\\
\hspace*{4cm}.\\
\hspace*{4cm}.\\
\hspace*{4cm}.\\
\textbf{For partition 5:} Samples 5, 10, 15, ... have been used as validation samples and the remaining as learning samples.

After partitioning data into learning set and validation set, the model selection for $\epsilon$ and $\gamma$ is performed using 5-fold cross-validation process. In this paper, to perform model selection, the $\epsilon$ and $\gamma$ values are varied over a range. The $\gamma$ value ranges from $2^{-7}$ to $2^{7}$ and $\epsilon$ value ranges from 0 to 5. Hence, ninty number of models will be generated to perform model selection operation.
\begin{table}[ht!]
  \centering
  \caption{Validation Errors Obtained Using SVR Linear Kernel for UCP}
  \renewcommand{\arraystretch}{1.4}
  \resizebox{7cm}{!}{
    \begin{tabular}{|c|c|c|c|c|c|c|}
    \hline
          & \textbf{$\epsilon$ = 0} & \textbf{1} & \textbf{2} & \textbf{3} & \textbf{4} & \textbf{5} \\
    \hline
    \textbf{$\mathbf{\gamma = 2^{-7}}$} & \textit{0.0009} & 0.2054 & 0.2054 & 0.2054 & 0.2054 & 0.2054 \\ \hline 
    \textbf{$\mathbf{2^{-6}}$} & 0.0009 & 0.2054 & 0.2054 & 0.2054 & 0.2054 & 0.2054 \\ \hline
    \textbf{$\mathbf{2^{-5}}$} & 0.0009 & 0.2054 & 0.2054 & 0.2054 & 0.2054 & 0.2054 \\ \hline
    \textbf{$\mathbf{2^{-4}}$} & 0.0009 & 0.2054 & 0.2054 & 0.2054 & 0.2054 & 0.2054 \\ \hline
    \textbf{$\mathbf{2^{-3}}$} & 0.0009 & 0.2054 & 0.2054 & 0.2054 & 0.2054 & 0.2054 \\ \hline 
    \textbf{$\mathbf{2^{-2}}$} & 0.0009 & 0.2054 & 0.2054 & 0.2054 & 0.2054 & 0.2054 \\ \hline
    \textbf{$\mathbf{2^{-1}}$} & 0.0009 & 0.2054 & 0.2054 & 0.2054 & 0.2054 & 0.2054 \\ \hline
    \textbf{$\mathbf{2^{0}}$} & 0.0009 & 0.2054 & 0.2054 & 0.2054 & 0.2054 & 0.2054 \\ \hline
    \textbf{$\mathbf{2^{1}}$} & 0.0009 & 0.2054 & 0.2054 & 0.2054 & 0.2054 & 0.2054 \\ \hline
    \textbf{$\mathbf{2^{2}}$} & 0.0009 & 0.2054 & 0.2054 & 0.2054 & 0.2054 & 0.2054 \\ \hline
    \textbf{$\mathbf{2^{3}}$} & 0.0009 & 0.2054 & 0.2054 & 0.2054 & 0.2054 & 0.2054 \\ \hline
    \textbf{$\mathbf{2^{4}}$} & 0.0009 & 0.2054 & 0.2054 & 0.2054 & 0.2054 & 0.2054 \\ \hline
    \textbf{$\mathbf{2^{5}}$} & 0.0009 & 0.2054 & 0.2054 & 0.2054 & 0.2054 & 0.2054 \\ \hline
    \textbf{$\mathbf{2^{6}}$} & 0.0009 & 0.2054 & 0.2054 & 0.2054 & 0.2054 & 0.2054 \\ \hline
    \textbf{$\mathbf{2^{7}}$} & 0.0009 & 0.2054 & 0.2054 & 0.2054 & 0.2054 & 0.2054 \\ 
    \hline
    \end{tabular}%
    }
  \label{tab:minvalderrlnrcp1}%
\end{table}%
\begin{table}[ht!]
  \centering
  \caption{Validation Errors Obtained Using SVR Polynomial Kernel for UCP}
  \renewcommand{\arraystretch}{1.4}
  \resizebox{7cm}{!}{
    \begin{tabular}{|c|c|c|c|c|c|c|}
    \hline
          & \textbf{$\epsilon$ = 0} & \textbf{1} & \textbf{2} & \textbf{3} & \textbf{4} & \textbf{5} \\
    \hline
    \textbf{$\mathbf{\gamma = 2^{-7}}$} & 0.0434 & 0.2054 & 0.2054 & 0.2054 & 0.2054 & 0.2054 \\ \hline 
    \textbf{$\mathbf{2^{-6}}$} & 0.0434 & 0.2054 & 0.2054 & 0.2054 & 0.2054 & 0.2054 \\ \hline
    \textbf{$\mathbf{2^{-5}}$} & 0.0434 & 0.2054 & 0.2054 & 0.2054 & 0.2054 & 0.2054 \\ \hline
    \textbf{$\mathbf{2^{-4}}$} & 0.0434 & 0.2054 & 0.2054 & 0.2054 & 0.2054 & 0.2054 \\ \hline
    \textbf{$\mathbf{2^{-3}}$} & 0.0431 & 0.2054 & 0.2054 & 0.2054 & 0.2054 & 0.2054 \\ \hline
    \textbf{$\mathbf{2^{-2}}$} & 0.0409 & 0.2054 & 0.2054 & 0.2054 & 0.2054 & 0.2054 \\ \hline
    \textbf{$\mathbf{2^{-1}}$} & 0.0257 & 0.2054 & 0.2054 & 0.2054 & 0.2054 & 0.2054 \\ \hline
    \textbf{$\mathbf{2^{0}}$} & 0.0059 & 0.2054 & 0.2054 & 0.2054 & 0.2054 & 0.2054 \\ \hline
    \textbf{$\mathbf{2^{1}}$} & 0.0058 & 0.2054 & 0.2054 & 0.2054 & 0.2054 & 0.2054 \\ \hline
    \textbf{$\mathbf{2^{2}}$} & 0.0058 & 0.2054 & 0.2054 & 0.2054 & 0.2054 & 0.2054 \\ \hline
    \textbf{$\mathbf{2^{3}}$} & 0.0058 & 0.2054 & 0.2054 & 0.2054 & 0.2054 & 0.2054 \\ \hline
    \textbf{$\mathbf{2^{4}}$} & 0.0058 & 0.2054 & 0.2054 & 0.2054 & 0.2054 & 0.2054 \\ \hline
    \textbf{$\mathbf{2^{5}}$} & 0.0058 & 0.2054 & 0.2054 & 0.2054 & 0.2054 & 0.2054 \\ \hline
    \textbf{$\mathbf{2^{6}}$} & 0.0058 & 0.2054 & 0.2054 & 0.2054 & 0.2054 & 0.2054 \\ \hline
    \textbf{$\mathbf{2^{7}}$} & \textit{0.0057} & 0.2054 & 0.2054 & 0.2054 & 0.2054 & 0.2054 \\ 
    \hline
    \end{tabular}%
    }
  \label{tab:minvalderrpolycp1}%
\end{table}%

Table \ref{tab:minvalderrlnrcp1} and \ref{tab:minvalderrpolycp1} show the validation error of ninty numbers of models generated for CP1 using SVR linear kernel and SVR polynomial kernel respectively based on the value of $\epsilon$ and $\gamma$. For SVR Linear kernel, \textit{0.0009} value has been chosen as the minimum validation error. Hence based on the minimum validation error, the best model is C = 0.99891, $\gamma = 2^{-7} (0.0078)$ and $\epsilon = 0$. Similarly for SVR Polynomial kernel, \textit{0.0057} value has been chosen as the minimum validation error. Hence based on the minimum validation error, the best model is C = 0.99891, $\gamma = 2^{7}(128)$ and $\epsilon = 0$.   

\begin{table}[ht!]
  \centering
  \caption{Validation Errors Obtained Using SVR RBF Kernel for UCP}
  \renewcommand{\arraystretch}{1.4}
  \resizebox{7cm}{!}{
    \begin{tabular}{|c|c|c|c|c|c|c|}
    \hline
          & \textbf{$\epsilon$ = 0} & \textbf{1} & \textbf{2} & \textbf{3} & \textbf{4} & \textbf{5} \\
    \hline
    \textbf{$\mathbf{\gamma = 2^{-7}}$} & 0.0370 & 0.2054 & 0.2054 & 0.2054 & 0.2054 & 0.2054 \\ \hline 
    \textbf{$\mathbf{2^{-6}}$} & 0.0312 & 0.2054 & 0.2054 & 0.2054 & 0.2054 & 0.2054 \\ \hline
    \textbf{$\mathbf{2^{-5}}$} & 0.0213 & 0.2054 & 0.2054 & 0.2054 & 0.2054 & 0.2054 \\ \hline
    \textbf{$\mathbf{2^{-4}}$} & 0.0105 & 0.2054 & 0.2054 & 0.2054 & 0.2054 & 0.2054 \\ \hline
    \textbf{$\mathbf{2^{-3}}$} & 0.0044 & 0.2054 & 0.2054 & 0.2054 & 0.2054 & 0.2054 \\ \hline 
    \textbf{$\mathbf{2^{-2}}$} & 0.0008 & 0.2054 & 0.2054 & 0.2054 & 0.2054 & 0.2054 \\ \hline
    $\mathbf{2^{-1}}$ & 0.0008 & 0.2054 & 0.2054 & 0.2054 & 0.2054 & 0.2054 \\ \hline
    \textbf{$\mathbf{2^{0}}$} & \textit{0.0007} & 0.2054 & 0.2054 & 0.2054 & 0.2054 & 0.2054 \\ \hline
    \textbf{$\mathbf{2^{1}}$} & 0.0007 & 0.2054 & 0.2054 & 0.2054 & 0.2054 & 0.2054 \\ \hline
    \textbf{$\mathbf{2^{2}}$} & 0.0010 & 0.2054 & 0.2054 & 0.2054 & 0.2054 & 0.2054 \\ \hline
    \textbf{$\mathbf{2^{3}}$} & 0.0011 & 0.2054 & 0.2054 & 0.2054 & 0.2054 & 0.0897 \\ \hline
    \textbf{$\mathbf{2^{4}}$} & 0.0012 & 0.2054 & 0.2054 & 0.2054 & 0.2054 & 0.2054 \\ \hline
    \textbf{$\mathbf{2^{5}}$} & 0.0016 & 0.2054 & 0.2054 & 0.2054 & 0.2054 & 0.2054 \\ \hline
    \textbf{$\mathbf{2^{6}}$} & 0.0020 & 0.2054 & 0.2054 & 0.2054 & 0.2054 & 0.2054 \\ \hline
    \textbf{$\mathbf{2^{7}}$} & 0.0020 & 0.2054 & 0.2054 & 0.2054 & 0.2054 & 0.2054 \\ 
    \hline
    \end{tabular}%
    }
  \label{tab:minvalderrrbfcp1}%
\end{table}%
\begin{table}[ht!]
  \centering
  \caption{Validation Errors Obtained Using SVR Sigmoid Kernel for UCP}
  \renewcommand{\arraystretch}{1.4}
  \resizebox{7cm}{!}{
    \begin{tabular}{|c|c|c|c|c|c|c|}
    \hline
          & \textbf{$\epsilon$ = 0} & \textbf{1} & \textbf{2} & \textbf{3} & \textbf{4} & \textbf{5} \\
    \hline
    \textbf{$\mathbf{\gamma = 2^{-7}}$} & 0.0401 & 0.2054 & 0.2054 & 0.2054 & 0.2054 & 0.2054 \\ \hline 
    \textbf{$\mathbf{2^{-6}}$} & 0.0370 & 0.2054 & 0.2054 & 0.2054 & 0.2054 & 0.2054 \\ \hline
    \textbf{$\mathbf{2^{-5}}$} & 0.0311 & 0.2054 & 0.2054 & 0.2054 & 0.2054 & 0.2054 \\ \hline
    \textbf{$\mathbf{2^{-4}}$} & 0.0210 & 0.2054 & 0.2054 & 0.2054 & 0.2054 & 0.2054 \\ \hline
    \textbf{$\mathbf{2^{-3}}$} & 0.0099 & 0.2054 & 0.2054 & 0.2054 & 0.2054 & 0.2054 \\ \hline
    \textbf{$\mathbf{2^{-2}}$} & 0.0046 & 0.2054 & 0.2054 & 0.2054 & 0.2054 & 0.2054 \\ \hline
    \textbf{$\mathbf{2^{-1}}$} & \textit{0.0014} & 0.2054 & 0.2054 & 0.2054 & 0.2054 & 0.2054 \\ \hline
    \textbf{$\mathbf{2^{0}}$} & 0.0054 & 0.2054 & 0.2054 & 0.2054 & 0.2054 & 0.2054 \\ \hline
    \textbf{$\mathbf{2^{1}}$} & 0.1014 & 0.2054 & 0.2054 & 0.2054 & 0.2054 & 0.2054 \\ \hline
    \textbf{$\mathbf{2^{2}}$} & 0.6041 & 0.2054 & 0.2054 & 0.2054 & 0.2054 & 0.2054 \\ \hline
    \textbf{$\mathbf{2^{3}}$} & 2.0331 & 0.2054 & 0.2054 & 0.2054 & 0.2054 & 0.2054 \\ \hline
    \textbf{$\mathbf{2^{4}}$} & 5.0454 & 0.2054 & 0.2054 & 0.2054 & 0.2054 & 0.2054 \\ \hline
    \textbf{$\mathbf{2^{5}}$} & 7.5381 & 0.2054 & 0.2054 & 0.2054 & 0.2054 & 0.2054 \\ \hline
    \textbf{$\mathbf{2^{6}}$} & 10.1991 & 0.2054 & 0.2054 & 0.2054 & 0.2054 & 0.2054 \\ \hline
    \textbf{$\mathbf{2^{7}}$} & 11.0733 & 0.2054 & 0.2054 & 0.2054 & 0.2054 & 0.2054 \\ 
    \hline
    \end{tabular}%
    }
  \label{tab:minvalderrsigcp1}%
\end{table}%

Similarly, Table \ref{tab:minvalderrrbfcp1} and \ref{tab:minvalderrsigcp1} show the validation error of ninty numbers of models generated for CP1 using SVR RBF kernel and SVR Sigmoid kernel respectively based on the value of $\epsilon$ and $\gamma$. For VR RBF kernel, \textit{0.0007} value has been chosen as the minimum validation error. Hence based on the minimum validation error, the best model is C = 0.99891, $\gamma = 2^{0}(1)$ and $\epsilon = 0$. Similarly for SVR Sigmoid kernel, \textit{0.0014} value has been chosen as the minimum validation error. Hence based on the minimum validation error, the best model is C = 0.99891, $\gamma = 2^{-1}(0.5)$ and $\epsilon = 0$. 

Finally based on model parameters value, the model has been again trained and tested using training and testing data set respectively to estimate the effort. 

\subsection{Performance Measures}
 The performance of the various models can be evaluated by using the following criteria \cite{Menzies}:
\begin{itemize}
\itemsep0em
\item The \textbf{Mean Square Error (MSE)} measures the average of the squares of the errors and is calculated as: 
\begin{equation} \label{msecomp}
MSE=\frac{\sum_{i=1}^{N}\left ( y_{i}-\bar{y} \right )^{2}}{N}
\end{equation}
where\\
\hspace*{0.1cm}$y_{i}$ = Actual Effort of $i^{th}$ test data.\\
\hspace*{0.1cm}$\bar{y}$ = Predicted Effort of $i^{th}$ test data.\\
\hspace*{0.1cm}N = Total number of data in the test set.
\item The \textbf{Mean Magnitude of Relative Error (MMRE)} can be achieved through the summation of MRE over N observations
\begin{equation} \label{mmrecomp}
MMRE=\sum_{1}^{N}\frac{\left |y_{i}-\bar{y}\right |}{y_{i}}
\end{equation}
\item The \textbf{Root Mean Square Error (RMSE)} is just the square root of the mean square error.
\begin{equation} \label{rmsecomp}
RMSE=\sqrt{MSE}
\end{equation}
\item The \textbf{Normalized Root Mean Square(NRMS)} can be calculated by dividing the RMSE value with standard deviation of the actual effort value for training data set.
\begin{equation} \label{nrmscomp}
NRMS=\frac{RMSE}{std(Y)}
\end{equation}
where Y is the actual effort for testing data set.
\item The \textbf{Prediction Accuracy (PRED)} can be calculated as:
\begin{equation} \label{predcomp}
PRED = (1 - (\frac{\sum_{i=1}^{N} \left | y_{i}-{\bar{y}} \right |}{N})) * 100
\end{equation}
\end{itemize}
While using the MMRE and PRED in evaluation, good results are implied by lower values of MMRE and higher value of PRED. After implementing the support vector regression based model using four different kernel methods for software effort estimation, the following results have been generated.\\[0.05cm]
\underline{\textbf{SVR Linear Kernel Result for UCP:}}\\[0.05cm]
\textbf{Param:  -s 3 -t 0 -c 0.9989 -g 0.0078 -p 0}\\[0.05cm]
* Mean Squared Error (MSE\_TEST) = 0.0026 (regression)\\[0.05cm]
* Squared correlation coefficient = 0.9932 (regression)\\[0.05cm]
* NRMS\_Test = 0.2431 \\[0.05cm]
\underline{\textbf{SVR Polynomial Kernel Result for UCP:}}\\[0.05cm]
\textbf{Param:  -s 3 -t 1 -c 0.9989 -g 128 -p 0}\\[0.05cm]
* Mean Squared Error (MSE\_TEST) = 0.0018 (regression)\\[0.05cm]
* Squared correlation coefficient = 0.9003 (regression)\\[0.05cm]
* NRMS\_Test = 0.4588 \\[0.05cm]
\underline{\textbf{SVR RBF Kernel Result for UCP:}}\\[0.05cm]
\textbf{Param:  -s 3 -t 2 -c 0.9989 -g 1 -p 0}\\[0.05cm]
* Mean Squared Error (MSE\_TEST) = 0.0030 (regression)\\[0.05cm]
* Squared correlation coefficient = 0.9881 (regression)\\[0.05cm]
* NRMS\_Test = 0.2822 \\[0.05cm]
\underline{\textbf{SVR Sigmoid Kernel Result for UCP:}}\\[0.05cm]
\textbf{Param:  -s 3 -t 3 -c 0.9989 -g 0.5 -p 0}\\[0.05cm]
* Mean Squared Error (MSE\_TEST) = 0.0034 (regression)\\[0.05cm]
* Squared correlation coefficient = 0.9920 (regression)\\[0.05cm]
* NRMS\_Test = 0.2764 \\[0.05cm]

The \textit{squared correlation coefficient}($r^{2}$) also known as the \textit{coefficient of determination} is one of the best approaches for evaluating the strength of a relationship. It is the proportion of variance in actual effort that can be accounted for by knowing use case point value for training data set. In the output generated, it is quite clearly mentioned that the \textit{squared correlation coefficient} value for different kernels is very high (greater than 0.9). Hence it can be concluded that there is a strong positive correlation between the use case point and the predicted effort required to develop the software i.e., a minor change in the use case point value results in significant change in the predicted effort value.

\begin{figure}[ht!]
\centering
\includegraphics[height=4.7cm, width=8cm]{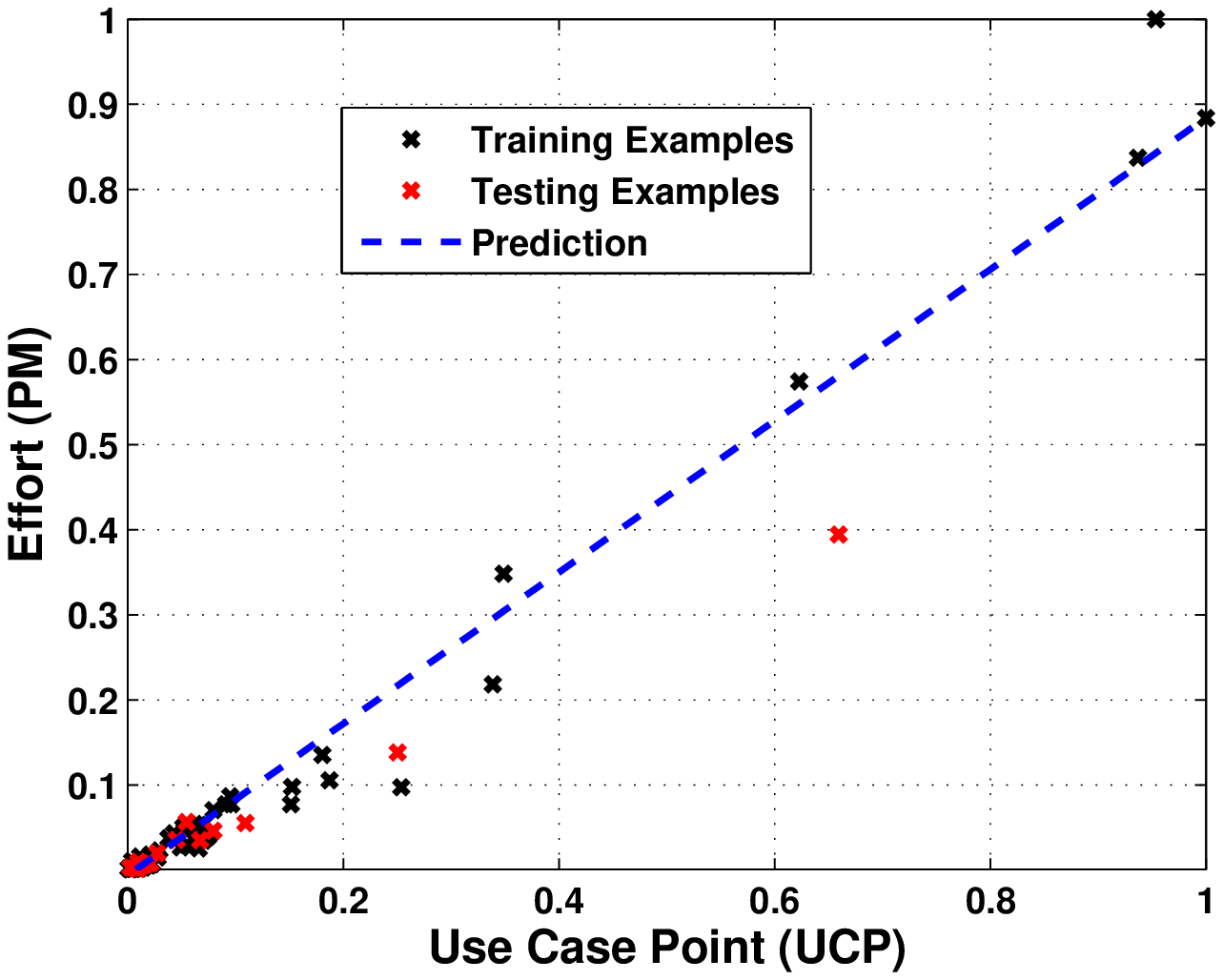}
\caption{SVR Linear Kernel based Effort Estimation Model for UCP}
\label{fig:svreffortlinear}
\end{figure}
\begin{figure}[ht!]
\centering
\includegraphics[height=4.7cm, width=8cm]{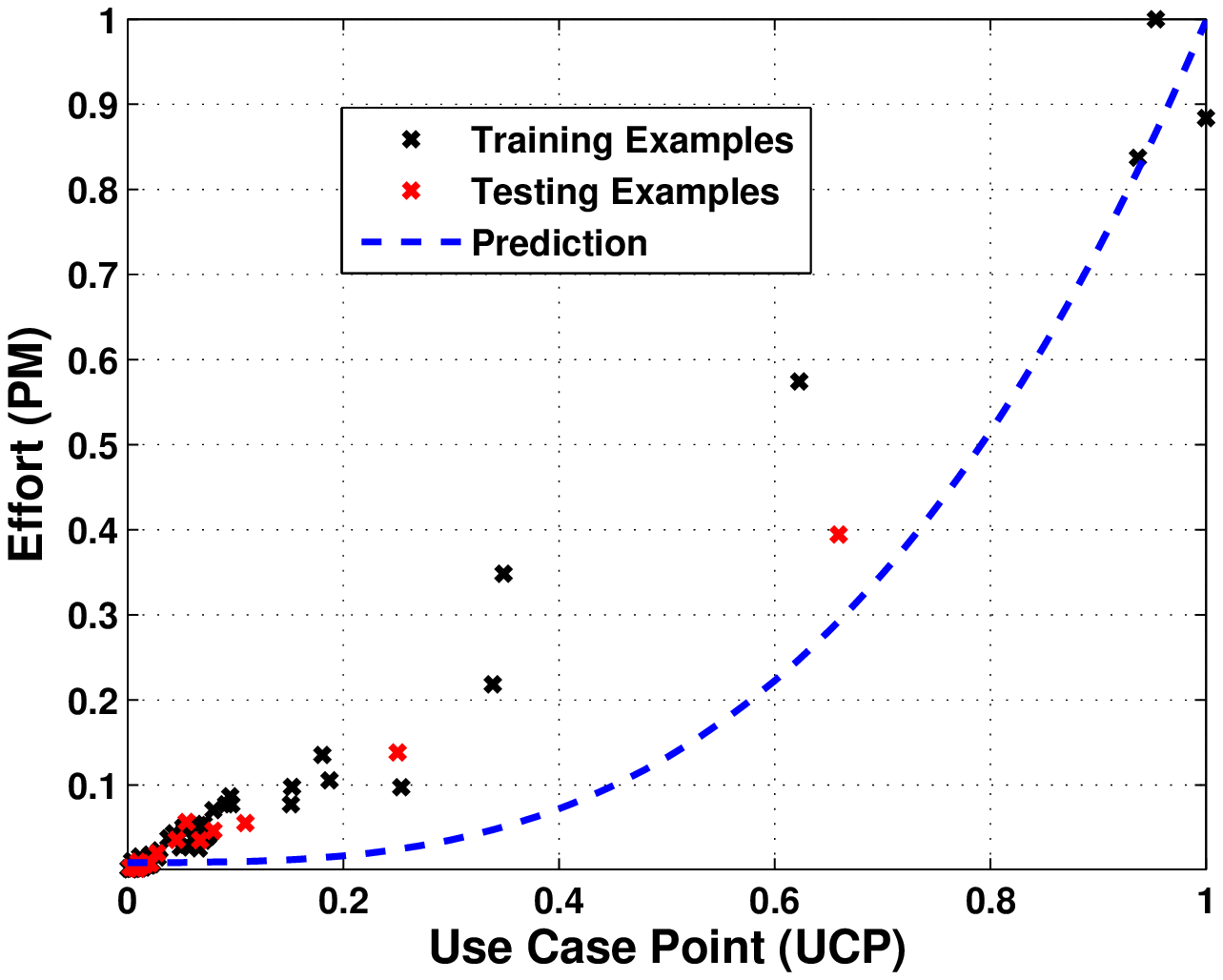}
\caption{SVR Linear Kernel based Effort Estimation Model for UCP}
\label{fig:svreffortpoly}
\end{figure}
\begin{figure}[ht!]
\centering
\includegraphics[height=4.7cm, width=8cm]{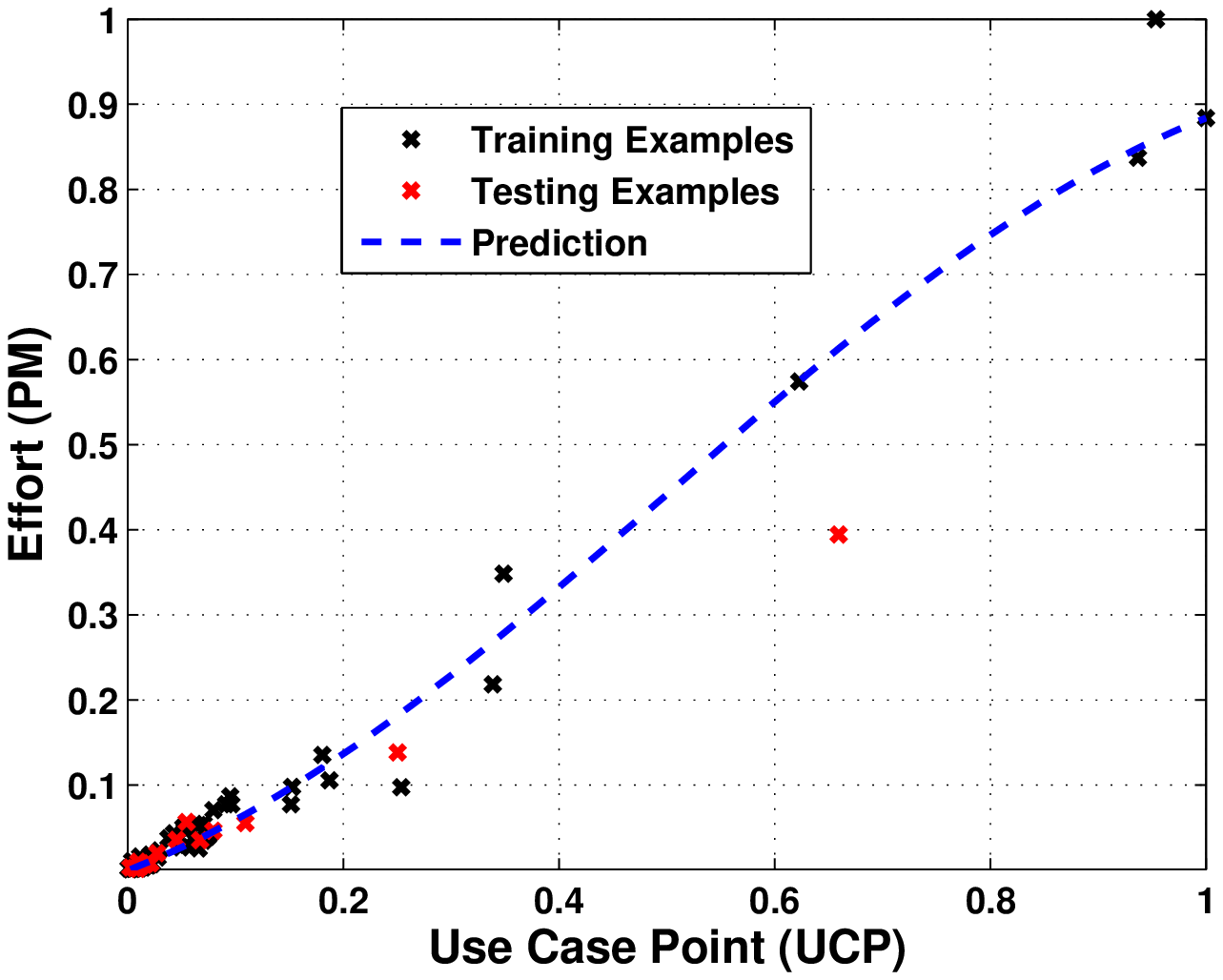}
\caption{SVR Linear Kernel based Effort Estimation Model for UCP}
\label{fig:svreffortrbf}
\end{figure}
\begin{figure}[ht!]
\centering
\includegraphics[height=4.7cm, width=8cm]{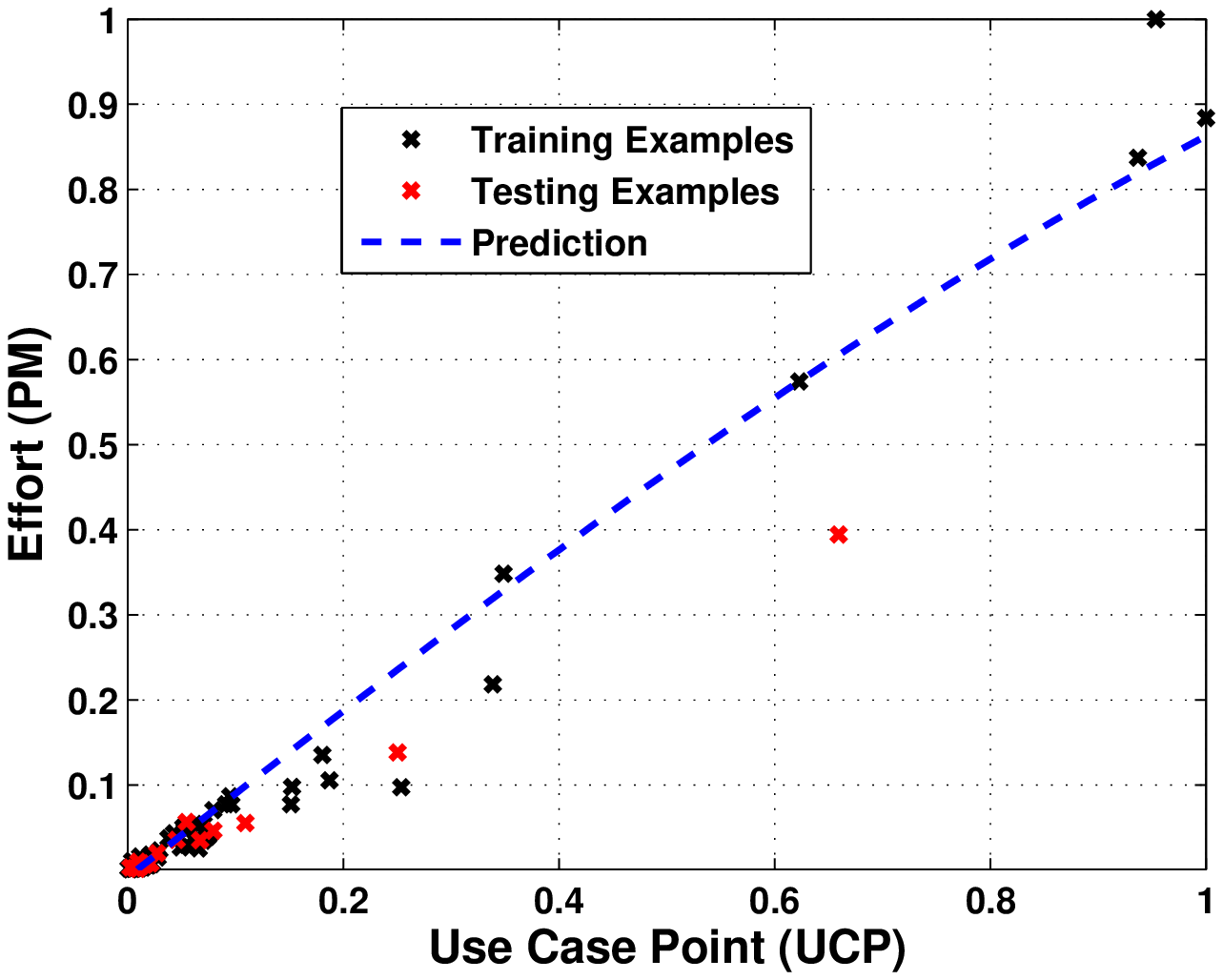}
\caption{SVR Linear Kernel based Effort Estimation Model for UCP}
\label{fig:svreffortsig}
\end{figure}
The proposed model generated using the SVR linear, polynomial, RBF and sigmoid kernel for UCP have been plotted based on the training and testing sample data set as shown in Figure \ref{fig:svreffortlinear}, \ref{fig:svreffortpoly}, \ref{fig:svreffortrbf} and \ref{fig:svreffortsig}. The graphs show the variation of predicted effort value with respect to its corresponding use case point value. In these graphs, it is clearly shown that the data points are very little dispersed than the regression line. Hence the correlation is higher. While comparing the dispersion of data points from the predicted model in the above graphs for UCP, it is clearly visible that for SVR RBF kernel based UCP model, the data points are less dispersed than other models. Hence, the models exhibit less error values and higher prediction accuracy values.

\section{COMPARISON}

On the basis of results obtained, the estimated effort using various SVR kernel methods are compared. The result shows that effort estimation using SVR RBF kernel based model gives less values of MMRE and higher prediction accuracy values than those obtained using other SVR kernel methods.

\begin{table}[htb!]
  \centering
  \caption{Comparison of Efforts obatined using various SVR kernel methods for UCP}
  \renewcommand{\arraystretch}{1.4}
  \resizebox{7cm}{!}{
  \large
    \begin{tabular}{|c|c|c|c|c|c|}
    \hline
    \textbf{} & \textbf{Actual Effort} & \textbf{SVR Linear Effort} & \textbf{SVR Polynomial Effort} & \textbf{SVR RBF Effort} & \textbf{SVR Sigmoid Effort} \\
    \hline
    \textbf{1} & \textbf{0.0100} & \textbf{0.0030} & \textbf{0.0090} & \textbf{0.0047} & \textbf{0.0018} \\ \hline
    \textbf{2} & \textbf{0.0031} & \textbf{-0.0035} & \textbf{0.0090} & \textbf{0.0010} & \textbf{-0.0054} \\ \hline
    \textbf{3} & \textbf{0.0024} & \textbf{-0.0003} & \textbf{0.0090} & \textbf{0.0028} & \textbf{-0.0018} \\ \hline
    \textbf{4} & \textbf{0.0015} & \textbf{0.0027} & \textbf{0.0090} & \textbf{0.0045} & \textbf{0.0015} \\ \hline
    \textbf{5} & \textbf{0.0061} & \textbf{0.0042} & \textbf{0.0090} & \textbf{0.0054} & \textbf{0.0031} \\ \hline
    \textbf{6} & \textbf{0.0050} & \textbf{0.0056} & \textbf{0.0090} & \textbf{0.0063} & \textbf{0.0047} \\ \hline
    \textbf{7} & \textbf{0.0064} & \textbf{0.0086} & \textbf{0.0090} & \textbf{0.0080} & \textbf{0.0080} \\ \hline
    \textbf{8} & \textbf{0.0083} & \textbf{0.1902} & \textbf{0.0090} & \textbf{0.0089} & \textbf{0.0096} \\ \hline
    \textbf{9} & \textbf{0.0082} & \textbf{0.1791} & \textbf{0.0090} & \textbf{0.0107} & \textbf{0.0129} \\ \hline
    \textbf{10} & \textbf{0.0200} & \textbf{0.1791} & \textbf{0.0090} & \textbf{0.0141} & \textbf{0.0191} \\ \hline
    \textbf{11} & \textbf{0.0356} & \textbf{0.1791} & \textbf{0.0091} & \textbf{0.0243} & \textbf{0.0366} \\ \hline
    \textbf{12} & \textbf{0.0568} & \textbf{0.1791} & \textbf{0.0092} & \textbf{0.0299} & \textbf{0.0457} \\ \hline
    \textbf{13} & \textbf{0.0352} & \textbf{0.1791} & \textbf{0.0093} & \textbf{0.0377} & \textbf{0.0581} \\ \hline
    \textbf{14} & \textbf{0.0464} & \textbf{0.1791} & \textbf{0.0095} & \textbf{0.0455} & \textbf{0.0701} \\ \hline
    \textbf{15} & \textbf{0.0553} & \textbf{0.1791} & \textbf{0.0103} & \textbf{0.0656} & \textbf{0.0990} \\ \hline
    \textbf{16} & \textbf{0.1383} & \textbf{0.1791} & \textbf{0.0246} & \textbf{0.1818} & \textbf{0.2357} \\ \hline
    \textbf{17} & \textbf{0.3945} & \textbf{0.1791} & \textbf{0.2926} & \textbf{0.6128} & \textbf{0.6052} \\ \hline
    \end{tabular}%
    }
  \label{tab:effortcompcp2}%
\end{table}%

Table.~\ref{tab:effortcompcp2} shows the comparison of actual effort with an estimated effort by various SVR kernel methods for UCP on the seventeen data taken for testing out of a data set of eighty four data.

    \begin{table}[htb!]
      \centering
       \large
      \caption{Comparison of Prediction Accuracy Values of Related Works}
      \label{relworkcomp}%
      \renewcommand{\arraystretch}{1.4}
      \resizebox{7cm}{!}{
        \begin{tabular}{|>{\centering}m{5cm}|>{\centering}m{4cm}|c|}
        \hline
        \multicolumn{1}{|c|}{\textbf{Related Papers}} & \multicolumn{1}{c|}{\textbf{Technique Used}} & \multicolumn{1}{c|}{\textbf{Prediction Accuracy}} \\ \hline
        \textbf{A. Issha et al. \cite{Issa}} & \textbf{3 Novel UCP model} & \textbf{67\%} \\ \hline
        \textbf{Ali B. Nasif et al. \cite{Nassif2012Treeboost} }  & \textbf{TreeBoost Model} & \textbf{88\%} \\ \hline
        \textbf{Ali B. Nasif et al. \cite{Nassif2012ANN} }  & \textbf{Artificial Neural Network Model} &  \textbf{90.27\%} \\ \hline
        \textbf{Ali B. Nasif et al. \cite{Nassif2011Regression} }  & \textbf{Regression Model} &  \textbf{95.8\%} \\ \hline
        \end{tabular}%
        }
    \end{table}%

Table~\ref{relworkcomp} provides a comparative study of the results obtained by some articles mentioned in the related work section. The performance of techniques used in those articles have been compared by measuring their prediction accuracy (PRED) values. Result shows that, a maximum of 95\% prediction accuracy is achieved using regression analysis technique for UCP. Finally, the results obtained in related work section is compared with results of proposed approaches, which is shown in Table~\ref{errorcomp}. The results obtained using proposed technique shows improvement in the prediction accuracy value.

\begin{table}[htb!]
  \centering
  \caption{Comparison of errors and prediction accuracy values obtained using various SVR kernel methods for UCP}
      \label{errorcomp}
      \renewcommand{\arraystretch}{1.4}
      \resizebox{7cm}{!}{
    \begin{tabular}{|c|c|c|}
    \cline{2-3}
    \multicolumn{1}{c|}{} & \textbf{MMRE} & \textbf{PRED} \\ \hline
    \textbf{SVR Linear Kernel} & \textbf{0.5438} & \textbf{97.7421\%} \\ \hline
    \textbf{SVR Polynomial Kernel} & \textbf{1.0003} & \textbf{97.4089\%} \\ \hline
    \textbf{SVR RBF Kernel} & \textbf{0.3857} & \textbf{98.0188\%} \\ \hline
    \textbf{SVR Sigmoid Kernel} & \textbf{0.6049} & \textbf{97.3931\%} \\ 
    \hline
    \end{tabular}%
    }
\end{table}%

Table~\ref{errorcomp} displays the final comparison of MMRE and PRED values for different SVR kernel methods. While comparing the obtained results with the results provided in related work section i.e., in Table~\ref{relworkcomp}, it can be observed that  the obtained results from proposed models provide better prediction accuracy values than the results obtained from models given in related work section. Similarly, the results obtained from different proposed models show that effort estimation using SVR RBF kernel gives less values of MMRE and higher values of prediction accuracy than those obtained using other SVR kernel methods.  

\section{CONCLUSION}

The Use Case Point Approach is one of the different effort estimation models that can be used for effort estimation of softwares developed using object oriented methodology,  because it is simple, fast, accurate to a certain degree. In this paper, first the use case point approach is used to estimate the effort involved in developing a software product. Then the results obtained have been optimized using four different support vector regression kernel methods. At the end of the study, a comparative analysis of the generated results has been presented in order to assess their accuracy. While comparing the results with the results provided in the related work section, it can be concluded that the results obtained from proposed models outperform the results generated by models given in related work section. Similarly, while comparing the results obtained using various SVR kernel methods, it can be concluded that for UCP, RBF kernel based support vector regression technique outperformed other three kernel methods. The computations for above procedure have been implemented and membership functions generated using MATLAB. This can be further analyzed and proved if real data for Use Point Approach is available. This approach can also be extended by using some other soft computing techniques such as Particle Swarm Optimization (PSO) and Genetic Algorithm (GA).

\bibliographystyle{Unsrt}
\bibliography{instruction}

\noindent{\includegraphics[width=1in,height=2in,clip,keepaspectratio]{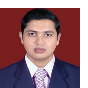}}
\begin{minipage}[b][1in][c]{1.8in}
{\centering{\bf {S M Satapathy}} did his MCA from
Silicon Institute of Technology and M.Tech in Computer Science \&
Engineering from KIIT University, Bhubaneswar, India. Currently he is pursuing his PhD}
\end{minipage}\\[0.1cm]
in Computer Science and Engineering at National Institute of Technology, Rourkela, India. His areas of interest are Software Cost Estimation, Program Slicing.\\\\
\noindent{\includegraphics[width=1in,height=2in,clip,keepaspectratio]{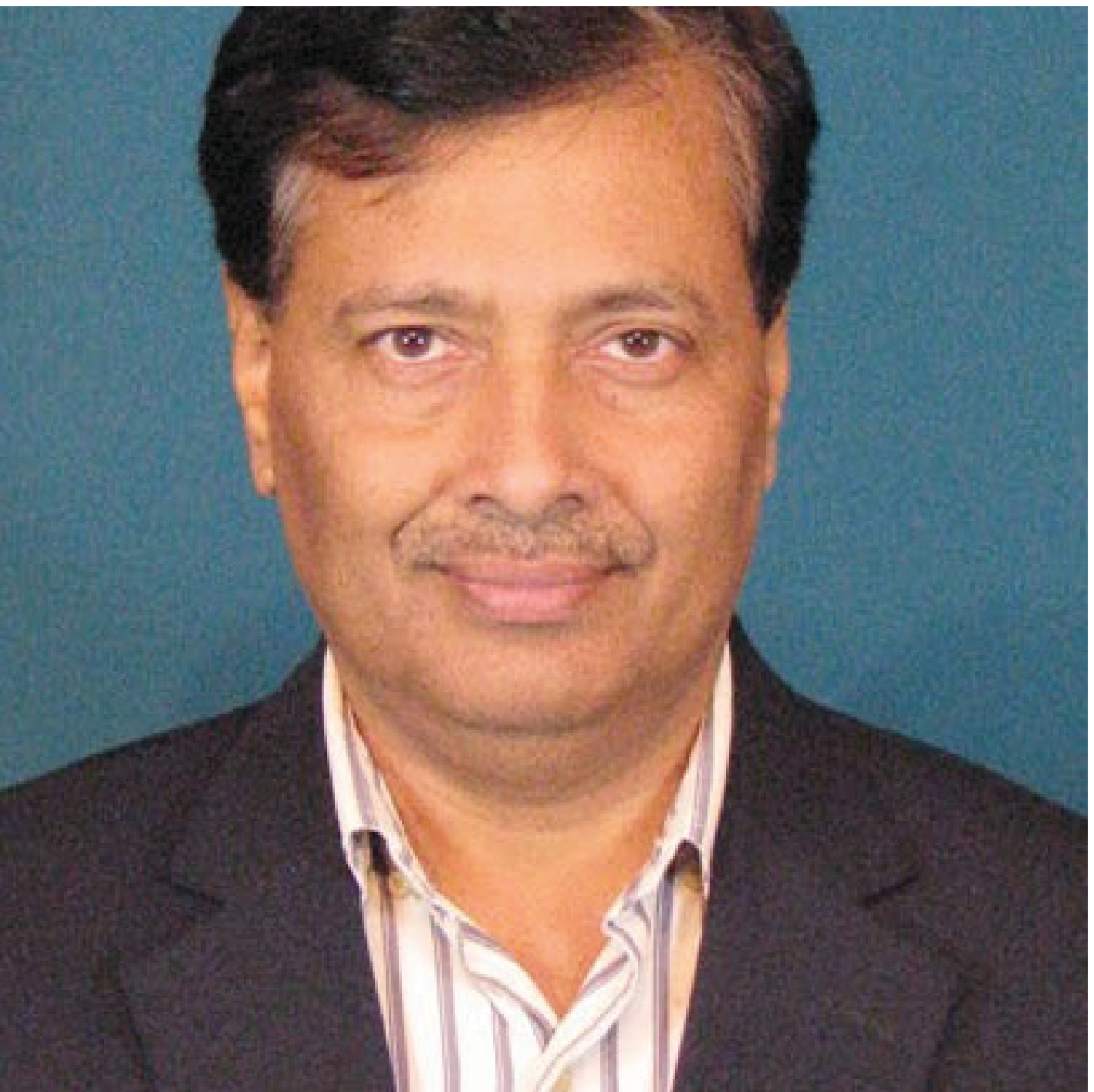}}
\begin{minipage}[b][1.1in][c]{1.74in}
{\centering{\bf{S K Rath }} is a Professor in the Department of
Computer Science and Engineering, NIT Rourkela since 1988. His
research interests are in Software Engineering, System Engineering, Bioinfo-}
\end{minipage}
rmatics \& Management. He has published a large number of papers in international journals and conferences in these areas. He is a Senior Member of the IEEE, USA and ACM, USA and Petri Net Society, Germany.

\end{document}